\documentclass[sigconf]{acmart}
\usepackage{algorithm,algorithmic,amsbsy,amssymb,booktabs,subfigure,url}

\usepackage{balance}

\def\BibTeX{{\rm B\kern-.05em{\sc i\kern-.025em b}\kern-.08emT\kern-.1667em\lower.7ex\hbox{E}\kern-.125emX}}

\copyrightyear{2019}
\acmYear{2019}
\setcopyright{acmcopyright}
\acmConference[MTD '19] {6th ACM Workshop on Moving Target Defense}{November 11, 2019}{London, United Kingdom}
\acmBooktitle{6th ACM Workshop on Moving Target Defense (MTD '19), November 11, 2019, London, United Kingdom}
\acmPrice{15.00}
\acmDOI{10.1145/3338468.3356824}
\acmISBN{978-1-4503-6828-5/19/11}

\begin{document}

\fancyhead{}

\title{A Cost-effective Shuffling Method against DDoS Attacks using Moving Target Defense}

\author{Yuyang Zhou}
\orcid{0000-0001-8626-0468}
\affiliation{%
  \institution{Southeast University}
  \city{Nanjing}
  \country{China}
}
\email{yyzhou@njnet.edu.cn}

\author{Guang Cheng}
\affiliation{%
  \institution{Southeast University}
  \city{Nanjing}
  \country{China}
}
\email{gcheng@njnet.edu.cn}

\author{Shanqing Jiang}
\affiliation{%
  \institution{Southeast University}
  \city{Nanjing}
  \country{China}
}
\email{sqjiang@njnet.edu.cn}

\author{Ying Hu}
\affiliation{%
  \institution{Southeast University}
  \city{Nanjing}
  \country{China}
}
\email{yhu@njnet.edu.cn}

\author{Yuyu Zhao}
\affiliation{%
  \institution{Southeast University}
  \city{Nanjing}
  \country{China}
}
\email{yyzhao@njnet.edu.cn}

\author{Zihan Chen}
\affiliation{%
  \institution{Southeast University}
  \city{Nanjing}
  \country{China}
}
\email{zhchen@njnet.edu.cn}

%
\renewcommand{\shortauthors}{Zhou and Cheng, et al.}

%
\begin{abstract}
Moving Target Defense (MTD) has emerged as a newcomer into the asymmetric field of attack and defense, and shuffling-based MTD has been regarded as one of the most effective ways to mitigate DDoS attacks. However, previous work does not acknowledge that frequent shuffles would significantly intensify the overhead. MTD requires a quantitative measure to compare the cost and effectiveness of available adaptations and explore the best trade-off between them. In this paper, therefore, we propose a new cost-effective shuffling method against DDoS attacks using MTD. By exploiting Multi-Objective Markov Decision Processes to model the interaction between the attacker and the defender, and designing a cost-effective shuffling algorithm, we study the best trade-off between the effectiveness and cost of shuffling in a given shuffling scenario. Finally, simulation and experimentation on an experimental software defined network (SDN) indicate that our approach imposes an acceptable shuffling overload and is effective in mitigating DDoS attacks.
\end{abstract}

%
%


%

%

%
\maketitle

\section{Introduction}\label{section1}
Due to the static nature of a cyber system, an attacker can not only perform reconnaissance on the target cyber system (i.e., scan the attack surface of the target system for possible vulnerabilities), but also launch an attack at his chosen time point to exploit the discovered vulnerabilities~\cite{carvalho2014moving}. The traditional strategy to defend the cyber system is to detect the unique behaviors of the attack. However, this strategy relies on knowing the characteristics of attacks. It becomes inefficient and insufficient when facing more advanced attacks with unknown behavioral patterns, which is common in today's cyber attacks. To counterbalance the advantage of attackers by reconnaissance, Moving Target Defense (MTD)~\cite{cai2016moving,lei2018moving} has emerged as a good mitigation technique that alters the static nature of cyber systems. MTD regularly changes certain aspects of the system to decrease an attackers' understanding of the target system. Any discovered vulnerabilities may disappear after enough time has passed, thus reducing the chance of a successful exploit. Essentially, MTD can increase the attack cost/complexity, and decrease the likelihood of successful attacks~\cite{manadhata2004measuring}.

Due to the frequent shuffling of attack surfaces by MTD, it becomes far more difficult for an adversary to launch a successful attack. But, its frequent shuffling can also have negative effects on the protected system by reducing the quality of service (QoS) on top of the extra costs associated~\cite{zangeneh2018cost}. In addition, when a random move transfers the attack surface to a new surface, there's a possibility that the new surface is more vulnerable than the previous surface. Therefore, it is necessary to assess both the cost and the effectiveness of available shuffling methods to find a balance between the two.

In this paper, a new cost-effective shuffling method is proposed to resist DDoS attacks using MTD. First, we describe a threat model to characterize the behavior of the attackers and defenders. Then, in order to model the interaction between the attacker and the defender as a game, we exploit the Multi-Objective Markov Decision Processes~\cite{roijers2017multi} to model the state transition of a system. Moreover, we will discuss the game process, the definition of the game payoff, and the generation of the game strategy to guide the defender to analyze the impact on the shuffle by the payoff of strategy. After that, we will propose the shuffling scenario and present our cost-effective shuffling algorithm (CES). The goal of CES is to find the optimal strategy for a sequence of shuffling decisions, which can reach the best trade-off between the effectiveness and the cost of shuffling. Our simulation and experiment results have shown that CES can effectively shuffle with limited cost to the SDN and performs well in resisting DDoS attacks.

The remainder of this paper is organized as follows. We discuss the related work in Section \ref{section2} and propose the threat model in Section \ref{section3}. Model specification and detailed analysis of the game are presented in Section \ref{section4}. Description of the shuffling scenario and algorithm are given in Section \ref{section5}. The performance of our proposed method is evaluated via simulation and experiment in Section \ref{section6}. Finally, we conclude the paper in Section \ref{section7}.
\section{Related Work}\label{section2}
Existing research on MTD shuffling can be classified into random, event-based, and hybrid mutation. Early research on random mutations~\cite{pal2013managed,rahman2014moving,gillani2015agile,chang2018fast} stipulate that each move in a random mutation occurs after a set time interval where the interval could be random or periodic. However, the time interval would be the only information needed in this case. In contrast, the moves in event-based mutations~\cite{carroll2014analysis,crouse2015probabilistic,zhang2017network} require extra information such as security policies and alerts. Upon receiving an external stimulus, the attack surface would be modified in order to mitigate the event. Hybrid mutations offer a mixed approach with combine many aspects of random and event-based mutations. Several researchers have proposed hybrid MTD models, such as Kampanakis et al.~\cite{kampanakis2014sdn}, who proposed a kind of network-level MTD techniques consisting of a hybrid mutation engine based on SDN. Huang and Ghosh~\cite{huang2011introducing} also proposed a system of servers where those offline could rotate in to replace those online, either at certain intervals or through certain events.

Some research was proposed to evaluate MTD mechanisms by quantifying the changes on the attack surface and assessing the cost and effectiveness of the mutation~\cite{kansal2017ddos,alavizadeh2018evaluation}. In order to assess the effectiveness of MTD techniques, Hong and Kim~\cite{hong2015assessing} developed a hierarchical attack representative model which is rather more flexible and scalable than common attack graphs. Bopche and Mehtre~\cite{bopche2017graph} employed classical graph distance metrics such as maximum common subgraph (MCS) and graph edit distance (GED) to measure temporal changes in attack surface of dynamic networks. Hong et al.~\cite{hong2018dynamic} also incorporated MTD techniques into a temporal graph-based graphical security model and developed a new set of dynamic security metrics to assess and compare their effectiveness. Moreover, an evaluation model of MTD effectiveness based on system attack surface (SAS) was proposed by Xiong et al.~\cite{xiong2019effectiveness}, and Zhang et al.~\cite{zhang2019efficient} proposed an efficient strategy selection for MTD, where the analytic hierarchy process (AHP) was employed to quantify the factors affecting the attack and defense costs.

In addition, some researchers adopted game theory as a tool to model the interaction between the attacker and the defender and determine the selection of MTD moving strategy. Prakash and Wellman~\cite{prakash2015empirical} employed empirical and game theoretic techniques to examine the interaction between the attacker and the defender and demonstrated that the efficiency of MTD is sensitive to its detection capability. Although they realized that security alerts play an important role in effective move selection, the cost of the moves was ignored. Feng et al.~\cite{feng2017signaling} proposed a Bayesian Stackelberg game that models the joint migration and signaling strategies for the defender in the face of a strategic and rational attacker and demonstrated that MTD can be improved through strategic information disclosure. Markov Decision Process(MDP) based approach has been utilized to analyze and further select optimal policies by many researchers~\cite{miehling2015optimal,hu2017online,zheng2018markov}, while Lei et al.~\cite{lei2018incomplete} proposed a novel of incomplete information Markov game theoretic approach to strategy generation. Although the proposed model has been examined via theoretical analysis and numerical study, the effectiveness in real world is still uncertain.

\section{Threat Model}\label{section3}
In this section, we describe a threat model to characterize the behavior of attackers and moving target defense mechanism. We assume a threat model in which the adversary has some rational attack strategies and needs to explore the target before strategy execution. The adversary may also have multiple network resources to scan and probe, although they may not utilize all of it when attacking targets. We also assume that the defender might take advantage of some defense mechanism to prevent the target system from being compromised. This theoretical framework follows the state-of-the-art MTD model proposed by Lin et al.~\cite{lin2017cost}.

\subsection{Attacker Behavior}
A strategic and rational attacker, with the objective of attacking the confidentiality, integrity and availability (CIA) of the attack target, always needs to obtain some sensitive parameters about the defenders before launching a successful attack. To gain knowledge of the defenders, an attacker may take the time, computing, and monetary resources to explore the protected system. Once the attacker determines that he has obtained enough information about the defender, the attack will be launched with characteristics that are systematically decided by the current system state as well as the defense actions. The whole procedure including probing and launching the attack incurs significant cost. For example, the attack cost of launching a DDoS attack will be related to the resources consumed by previous IP address scanning, stealthy port scanning and the amount of utilized clients when the attack happens.

\subsection{Defense Mechanism}
In order to guard a system from being hacked or destroyed, the defender has to collect the information about the whole system and find any suspicious behavior that may lead to risks. Using moving target defenses to safeguard the system, the defender needs to make shuffles to change the attack surface as well as taking other necessary measures against an attacker. For each shuffle, it incurs a shuffle cost due to the utilized computing and network resources. In detail, this paper is focused on shuffle based MTD techniques that can be implemented at network level.

Therefore, the defense mechanism is defined as follows. Once one or several hosts in the protected system are compromised, in order to prevent the follow-up, the defender will shuffle the exploited hosts by the following defense types.

$\bullet$ Port hopping: Dynamic and continuous change of port number of a particular service.

$\bullet$ IP hopping: The defender changes the IP address of a VM dynamically and incessantly.

$\bullet$ Migration: The defender migrates the applications or services under attack between VMs.

\subsection{Objective}

The objective of this paper about cost-effective shuffling MTD method is to investigate the optimal way for a defender to make decisions while taking into account both the shuffling/attack cost and effectiveness between the defender and the attacker. It is important to maximize the shuffling effectiveness and minimize the cost while restricting the attacker's payoff and forcing them to terminate the attack. Moreover, it is possible for a defender to endure risks without shuffling, if the shuffling cost is high while the effectiveness is low. We seek to examine what is the best way to make the shuffling decision and how to reach the best trade-off between cost and effectiveness.

\section{Game Model}\label{section4}
Many real-world decision problems have multiple objectives. For example, for a
computer network we may want to maximize performance while minimizing power consumption~\cite{roijers2017multi}. The field of multi-objective decision making addresses how to formalize and solve decision problems with multiple objectives.

In the following, we exploit Multi-Objective Markov Decision Processes (MOMDP) to model the interaction between the attacker and the defender as a game, with the objective of maximizing the defender's payoff and minimizing the attacker's payoff. The shuffling selection process can be modeled as a sequential game in which the defender is the leader and the attacker is the follower. A MOMDP for two objectives in our case is a tuple $(t, S, O, A, D, R, C, \gamma)$, where:

$\bullet$ $t$ is the time step of a game, and $t \in \left\{0,...,T \right\}$ where $T$ is the time horizon.

$\bullet$ $S$ represents a finite set of states, including all possible attack surfaces that the protected system could experience and let $S_t$ be the state of the system at time step $t$.

$\bullet$ $O$ represents the status of services or VMs by defender's observation with confidence coefficient $\pi \in [0, 1]$.

$\bullet$ $A$ denotes a finite set of attacker actions, and let $A_t$ be the attacker's action at time step $t$.

$\bullet$ $D$ denotes a finite set of defender actions, and let $D_t$ be the defender's action at time step $t$.

$\bullet$ $R$: $S \times A(D)\times S \rightarrow R$ is a rewarding function that maps a state and an action to a reward for the player.

$\bullet$ $C$: $A(D) \rightarrow C$ assigns a cost to each action the players take.

$\bullet$ $\gamma$ is the discount factor where $\gamma \in (0, 1]$.

In this game, the defender adopts an MTD strategy by migrating the resource across the network to make it difficult for the attacker to identify the real location of the resource, while the attacker may observe the defender's actions by monitoring network traffic. Knowing this strategy (but not its realization), the attacker then determines against which VM to conduct DDoS attack and which IP address to choose. The defender can also obtain the state of the protected system and attacker's actions by observation. Thus, both will play their best strategy to act against their opponent.

\subsection{Game Process}\label{subsection4.1}
\begin{algorithm}[t]
\caption{State Transition Function}\label{alg1}
    \begin{algorithmic}[1]
    \REQUIRE ~~\\
    The system state at time step $t$, $S_t$;\\
    The observation by defender at time step $t$, $O_t$;\\
    The defender action at time step $t+1$, $D_{t+1}$;\\
    The attacker action at time step $t+1$, $A_{t+1}$;\\
    \ENSURE ~~\\
    The system state probability distribution at time step $t+1$, $S_{t+1}$ with probability $p$;\\
    \IF{$O_t(v) \subseteq S_t(v)$}
    \STATE $A_{t+1}(v) \gets 1$;
    \IF{$v \in D_{t+1}(v)$}
    \STATE $S_{t+1}(v) \gets 0$;
    \ELSE
    \STATE with probability $p(v)$, $S_{t+1}(v) \gets 1$;
    \ENDIF
    \ELSE
    \STATE $S_{t+1}(v) \gets S_{t}(v)$;
    \IF{$v \in D_{t+1}(v) $ and $ v \in A_{t+1}(v)$}
    \STATE $S_{t+1}(v) \gets 0$;
    \ELSE
    \IF{$v \notin D_{t+1}(v)$ and $ v \in A_{t+1}(v)$}
    \STATE with probability $p(v)$, $S_{t+1}(v) \gets 1$;
    \ELSE
    \FOR{$v\notin{D_{t+1}(v)}$ and ${v\notin{A_{t+1}(v)}}$}
    \STATE with probability $p(v',v)$, $S_{t+1}(v) \gets 1$;
    \ENDFOR
    \ENDIF
    \ENDIF
    \ENDIF
    \end{algorithmic}
\end{algorithm}

\noindent At the beginning of the game, $S_0$, $A_0$ and $D_0$ need to be initialized with $\varnothing$. Based on our assumption, the attacker is fully aware of system state at every time step, whereas the defender only knows the initial state $S_0$ and needs to observe the system to obtain the subsequent states. Thus, we also set $O_0 = \varnothing$.

At each time step $t+1 \in \left\{1,...,T \right\}$, the attacker can choose any VM $v \in V$ to conduct DDoS attack with a success probability $p(v)$, and $p(v,v')$ from VM $v$ to $v'$ if the attacker has taken control of {\it v}. Simultaneously, the defender decides which VMs to shuffle to prevent the attacker from further intruding.

After the initialization, the game proceeds in discrete time steps, $t+1 \in \left\{1,...,T \right\}$, with both players aware of the current time. The following sequence of game events between the defender and attacker occurs at each time step $t+1$.

(1)The attacker observes $S_t$, while the defender observes $O_t$.

(2)The attacker and defender select their actions $A_{t+1}$ and $D_{t+1}$ according to their respective strategies at the same time.

(3)The system transits to its next state $S_{t+1}$ according to the transition function(Algorithm \ref{alg1}).

(4)The attacker and defender evaluate their rewards and costs for the time step, respectively.

(5)The attacker and defender enter the next time step unless the time step $T$ has arrived.

\subsection{Game Payoff}
As discussed in Section~\ref{subsection4.1}, $S_t$ is a system state at time step $t$, when the attacker plays $A_t$, the defender plays $D_t$ and the previous system state is $S_{t-1}$. We denote by $H_T = \left\{(S_0, A_0, D_0), . . . , (S_T, A_T, D_T)\right\}$ the game history, which consists of all system states and players' actions at each time step.

After both players have taken actions in the game, each of them will get either a negative or a positive return. It is the quantitative assessment of each player's action which represents the game payoff. In MTD, both the attacker and the defender need to take the payoff into consideration when they make attack or defense decisions. Each player then receives a payoff function and aims to increase their own expected payoffs.

With respect to $H_T$, the defender and the attacker's payoff values of two objectives, which include goal rewards and action costs, can be separately presented as follows:

\begin{subequations}
\begin{equation}\label{equ1a}
P^d{(H_T)}\!=\!\sum\limits_{t\!+\!1\!=\!1}^T\!\gamma^t\!\left[\sum\limits_{S_{t\!+\!1}(v)=0}\!R^d(v)\!-\!\sum\limits_{v\in D_{t+1}}\!C^d(v)\right]
\end{equation}
\begin{equation}\label{equ1b}
\begin{aligned}
P^a{(H_T)}\!&=\!\sum\limits_{t\!+\!1\!=\!1}^T\!\gamma^t\!\left[\sum\limits_{S_{t\!+\!1}(v)=1}\!R^a(v)\!-\!\sum\limits_{v\in A_{t+1}}\!C^a(v)\right.\\
&\left.-\sum\limits_{v\notin D_{t+1} \cap v\notin A_{t+1}}\!C^a(v)\right]
\end{aligned}
\end{equation}
\end{subequations}

\subsection{Game Strategy}
As discussed above, both players will play their best strategies to act against the opponent and aim to maximize the value of payoff function $P$, which depends on the distribution of $H_T$. To analyze the game process more meticulously, heuristic strategies for both players are proposed in this section to depict detailed actions between them.

\subsubsection{Attacker Strategy}
For the attackers, at time step $t+1$, based on $S_t$ , they need to consider only VM $v \in V$ that can change the target system state at time step $t+1$. Hence, we denote by $\alpha(S_t)$  the $potential \ attack \ target$ at time step $t+1$ which represents this set of VMs and consists of two parts as follows:

(1)Target on VM $v$ directly to launch an attack.

(2)Target on another VM $v'$ with probability to reach $v$.

Based on the two parts of VMs discussed above, we obtain $\alpha(S_t)$ defined as follows:\\
\begin{equation}\label{equ2}
\begin{aligned}
\alpha(S_t)\!&=\!\left\{v\!\in\!V|S_t{(v)\!=\!0}\right\}\\
&\cup\left\{v'\!\in\!V|S_t{(v')\!=\!0,p(v'\!,\!v)\!>\!0}\right\}
\end{aligned}
\end{equation}

Since the attacker is rational in our assumption, the attacker chooses actions based on quantitative assessment of the game payoff with $\alpha(S_t)$. Intuitively, the value of an attack payoff quantitatively represents what the attacker can obtain by this attack at the time step.

The main idea of this game strategy is to choose the attack target by which the attacker's payoff could be maximized based on previous system state at each time step. However, due to lack of knowledge about the defender's action at this time step, the payoff the attacker calculates is biased for their unilateral action. This attack strategy generation is illustrated in Algorithm~\ref{alg2}.
   \begin{algorithm}[t]
  \caption{ Attack Strategy Generation}\label{alg2}
  \begin{algorithmic}[1]
    \REQUIRE ~~\\
    The system state at time step $t$, $S_t$;\\
    \ENSURE ~~\\
    The attacker action at time step $t+1$, $A_{t+1}$;\\
    \STATE Initialize $P^a{(v)} \gets 0$;\\
    \FOR{$S_t(v)=0,v \in V$}
    \STATE Calculate $R^a(v),C^a(v)$;\\
    \IF {$\gamma^t(p(v)R^a(v)-C^a(v))>P^a(v)$}
    \STATE $P^a(v)\gets \gamma^t(p(v)R^a(v)-C^a(v))$;\\
    \STATE Update $v \in P^a(v)$;\\
    \FOR{$S_t(v')=0,v' \in V$}
    \STATE Calculate $R^a(v'),C^a(v')$;\\
    \IF{$\gamma^t(p(v',v)R^a(v')-C^a(v'))\leq P^a(v)$}
    \STATE Retain $v \in P^a(v)$;\\
    \ELSE
    \STATE $P^a(v)\gets \gamma^t(p(v',v)R^a(v')-C^a(v'))$;\\
    \STATE Update $v \in P^a(v)$;\\
    \ENDIF
    \ENDFOR
    \ENDIF
    \ENDFOR
    \IF{$P^a(v) \leq 0$}
    \STATE $A_{t+1} \gets \varnothing$;
    \ELSE
    \STATE $A_{t+1} \gets \left\{v \in V|S_t(v)=0,v \in P^a(v)\right\}$;
    \ENDIF
    \RETURN $A_{t+1}$
  \end{algorithmic}
\end{algorithm}

   \begin{algorithm}[t]
  \caption{ Defend Strategy Generation}\label{alg3}
  \begin{algorithmic}[1]
    \REQUIRE ~~\\
    The observation by defender at time step $t$, $O_t$;\\
    The defender action at time step $t$, $D_t$;\\
    The number of VMs, $n$;\\
    \ENSURE ~~\\
    The defender action at time step $t+1$, $D_{t+1}$;\\
    \STATE Initialize $P^d{(v)} \gets 0$;\\
    \FOR{$O_t(v)=1,v \in V$}
    \STATE Calculate $R^d(v),C^d(v)$;\\
    \IF {$\gamma^t(\pi(v)R^d(v)-C^d(v))\leq P^d(v)$}
    \STATE Retain $v \in P^d(v)$;\\
    \ELSE
    \STATE $P^d(v)\gets \gamma^t(\pi(v)R^d(v)-C^d(v))$;\\
    \STATE Update $v \in P^d(v)$;\\
    \ENDIF
    \ENDFOR
    \FOR{$O_t(v)=0$ and $v \notin D_t(v)$}
    \STATE Calculate $R^d(v),C^d(v)$;\\
    \IF{$\gamma^t(\dfrac{\pi(v)R^d(v)}{n-1}-C^d(v)) \leq P^d(v)$}
    \STATE Retain $v \in P^d(v)$;\\
    \ELSE
    \STATE $P^d(v)\gets \gamma^t(\dfrac{\pi(v)R^d(v)}{n-1}-C^d(v))$;\\
    \STATE Update $v \in P^d(v)$;\\
    \ENDIF
    \ENDFOR
    \IF{$P^d(v) \leq 0$}
    \STATE $D_{t+1} \gets \varnothing$;
    \ELSE
    \STATE $D_{t+1} \gets \left\{v \in V \cap v \in P^d(v)\right\}$;
    \ENDIF
    \RETURN $D_{t+1}$
  \end{algorithmic}
\end{algorithm}
\subsubsection{Defender Strategy}
For the defenders, since they do not know the true system states at each time step, it is crucial for them to reason through the possible system states based on their observations before committing to a defensive action. As mentioned in the Section~\ref{subsection4.1}, in our game, the defender only knows the initial system state $S_0$, where $S_0(v)=0$ for each $v \in V$.

The defender needs to take both their observation and their assumptions about the attacker strategy into consideration to form an understanding of the current system state. Similarly, we denote by $\beta(O_t)$  the $potential$ $defend$ $target$ at time $t+1$ as follows:

(1)Target on VM $v$ according to defender's observation $O_t$.

(2)Target on VM which is not in $D_t$.

According to the above analysis, we obtain $\beta(O_t)$ defined as follows:
\begin{equation}\label{equ3}
\begin{aligned}
\beta(O_t)\!&=\!\left\{v\!\in\!V|O_t{(v)\!=\!1}\right\}\\
&\cup \left\{v\!\in\!V|O_t(v)\!=\!0\cap v\!\notin\!D_t(v)\right\}
\end{aligned}
\end{equation}

As a rational defender, before making decisions, he also needs to assess the game payoff of imminent actions with $\beta(O_t)$. The quantitative assessment of the game payoff for the defender represents the quality of the strategy to fight against attacker's malicious actions at that time step. Essentially, the higher the value of the defense payoff is, the safer the protected system will be. This defense strategy generation is illustrated in Algorithm~\ref{alg3}.

\section{Cost-Effective Shuffling Method}\label{section5}
As discussed above, we give the description of the game model and describe the game process and game strategies between both the attacker and the defender. However, the game may reach an equilibrium which is undesirable for the defender. To make the game more beneficial for the defender and reach the best trade-off between shuffling cost and defense effectiveness, we propose a cost-effective shuffling method, which consists of threat model and game theory, to adopt different shuffling types under different conditions.

\subsection{Shuffling Scenario}
When a service or a VM is under DDoS attacks, the defender controls and relocates the ports, IPs or VMs in use from extra resources. Nevertheless, additional overhead is incurred in the procedure of a shuffle. Therefore, our goal is to balance the defense effectiveness and the overhead whereas restricting the attacker's payoff by the implementation of a shuffling method.

To increase the applicability of our shuffling method and expound the details more clearly, we make some assumptions and propose the shuffling scenario as follows.

\textbf{Given}: a set of $q$ users and a group of $n$ VMs with $r$ network segments and $u$ ports of equal resources for $m$ users, where $m \times n = q, r \leqslant n$

\textbf{Output}:three sequences of matrices $(X_0,X_1,...,X_T)$,$(Y_0,Y_1,...,Y_T),\\(Z_0,Z_1,...,Z_T)$, where $X_t \in \left\{0,1\right\}^{r \times n}$,$Y_t \in \left\{0,1\right\}^{u \times n},Z_t \in \left\{0,1\right\}^{q \times n}$, such that
\begin{subequations}
\begin{equation}\label{equ4a}
\sum\limits_{i=1}^n x^t_{ij} \geqslant 1 \quad j=1,...,r;
\end{equation}
\begin{equation}\label{equ4b}
\sum\limits_{j=1}^r x^t_{ij} = 1 \quad i=1,...,n;
\end{equation}
\begin{equation}\label{equ4c}
\sum\limits_{i=1}^n y^t_{ij} \leqslant n \quad j=1,...,u;
\end{equation}
\begin{equation}\label{equ4d}
\sum\limits_{i=1}^n z^t_{ij} = 1 \quad j=1,...,q;
\end{equation}
\begin{equation}\label{equ4e}
\sum\limits_{j=1}^q z^t_{ij} = m \quad i=1,...,n;
\end{equation}
\end{subequations}

The matrix $X_t$ represent the IP shuffling decision at time step $t$, where binary variable $x^t_{ij}$ indicates that whether the $i$-th VM is assigned to $j$-th network segment. Hence, Equation~\ref{equ4a} states that each network segment owns at least one VM, and Equation~\ref{equ4b} ensures that each VM is assigned to only one network segment. Similarly, The matrix $Y_t$ represent the port shuffling decision at time step $t$, where binary variable $y^t_{ij}$ indicates that whether the $i$-th VM is assigned to $j$-th port. Hence, we can easily get Equation~\ref{equ4c} which indicates that at most $n$ VMs share the same port number.

As the VM migration is the third shuffling mechanism, the matrix $Z_t$ denote the overall condition of VM migration at time step $t$ and the binary variable $z^t_{ij}$ represents that whether the $j$-th user is assigned to $i$-th VM. Based on Equation~\ref{equ4d} and Equation~\ref{equ4e}, we can conclude that each user is assigned to only one VM and each VM can only be allowed to serve $m$ users.

\subsection{Cost-Effective Shuffling Algorithm}
In the following, we first present a cost-effective shuffling algorithm to consider the cost and effectiveness of shuffling, with the two objectives of maximizing the payoff that the defender may obtain and minimizing the payoff which the attacker can get.

Specifically, in the initial assignment step , $q$ users, $r$ network segments and $u$ ports are randomly assigned to $n$ VMs in our shuffling scenario, whereas the $t$-th shuffling step iteratively reduces the number of the crashed VMs. Afterwards, the system state at time step $t$ represents the assignment of users, network segments, ports in the system and the condition of crashed VMs through state transition function (Algorithm \ref{alg1}), which requires the defender's and attacker's strategies as the input. As discussed in Algorithm~\ref{alg2} and Algorithm~\ref{alg3}, the generation of strategies is directly related to the rewards and costs of their actions.

Hence, the defender's rewarding value $R^d_{t+1}$ and cost value $C^d_{t+1}$ at each time step $t$ with state transition function $STF$ represent the effectiveness and cost of a shuffle as follows:

\begin{subequations}
\begin{equation}\label{equ5a}
\begin{aligned}
R^d_{t+1}&=\sum\limits_{S_{t+1}(v)=0} R^d(v)\\
&=\sum\limits_{v\in V} STF \left(S_t(v)-S_{t+1}(v)\right)
\end{aligned}
\end{equation}
\begin{equation}\label{equ5b}
\begin{aligned}
C^d_{t+1}&=\sum\limits_{v \in D_{t+1}}C^d(v)=\sum\limits_{v\in D_{t+1}} (w_1\sum\limits_{j=1}^r|x_{vj}^{t+1}-x_{vj}^t|\\
&+w_2\sum\limits_{j=1}^u|y_{vj}^{t+1}-y_{vj}^t|+w_3\sum\limits_{j=1}^q|z_{vj}^{t+1}-z_{vj}^t|)
\end{aligned}
\end{equation}
\end{subequations}

Similarly, the attacker's rewarding value $R^a_{t+1}$ and cost value $C^a_{t+1}$ respectively represent the reward obtained from the VM crash and the cost caused during the whole attack stages, which can be calculated by follows:

\begin{subequations}
\begin{equation}\label{equ6a}
\begin{aligned}
R^a_{t+1}&=\sum\limits_{S_{t+1}(v)=1} R^a(v)\\
&=\sum\limits_{v\in V} STF \left(S_t(v)-S_{t+1}(v)\right)
\end{aligned}
\end{equation}

\begin{equation}\label{equ6b}
\begin{aligned}
C^a_{t+1}&=\sum\limits_{v \in D_{t+1}}C^a(v)+\sum\limits_{{v \notin D_{t+1}} \cap {v \notin A_{t+1}}}C^a(v)\\
&=\sum\limits_{v \in D_{t+1}}w_3+\sum\limits_{{v \notin D_{t+1}} \cap {v \notin A_{t+1}}}(w_1+w_2)
\end{aligned}
\end{equation}
\end{subequations}

Regarding the defender's shuffling effectiveness, the rewarding function in Equation~\ref{equ5a} represents the status transition from time step $t$ to $t+1$. In terms of IP hopping, port hopping and migration cost in a shuffle, the cost function in Equation~\ref{equ5b} represents the cost of shuffling from time step $t$ to $t+1$, where $w_1, w_2, w_3$ is the weights assigned by the network operator.

Instead, the reward function of the attacker in Equation~\ref{equ6a} indicates that the target VM has been crashed at time step $t+1$. Moreover, the attacker's cost value in Equation~\ref{equ6b} can be divided into two parts: the cost occurs when implementing the attack to the target VM and the cost spends during the scanning and probing stages. Hence, it can be calculated by the number of VMs, ports and IPs, which meet the strategies of the defender and attacker, using the assigned weights $w_1, w_2, w_3$ as well.

Thereout, we can obtain the payoff values of the defender and attacker across the whole game history, using Equation~\ref{equ1a}-\ref{equ1b}, Equation~\ref{equ5a}-\ref{equ5b} and Equation~\ref{equ6a}-\ref{equ6b} in our shuffling scenario. With the objective of maximizing the defender's payoff and minimizing the attacker's payoff, we utilize the difference between the payoff values to find the optimal trade-off between the effectiveness and cost among both players.

However, in an actual scenario, not all users are online at the same time and unnecessary shuffling costs are generated during each time step. In order to decrease the extra costs, we denote the number of online users in one VM by $\eta$ to guide the defender to make his decision in a more cost-effective manner, where $0 \leqslant \eta \leqslant m$. Thereout, CES (Algorithm~\ref{alg4}) aims to significantly reduce the unnecessary cost, restrict the attacker's payoff, and find the optimal shuffling decisions for the defender at each time step. In CES, Line 1 has a holistic view of the crashed VMs based on the current observation of the system. Then, Line 2 judges whether the current VM has online users and no shuffling decisions are given in Line 3 if there is no user. Moreover, if existing online users, a previous threshold has been set in Line 5. According to both players' payoff values and the average number of users in each VM, different shuffling decisions are made in Line 6-8 and Line 10-13 separately. Finally, shuffling decisions of all VMs for the next time step are returned in Line 17.

\begin{algorithm}[t]
  \caption{ Cost-Effective Shuffling Algorithm(CES)}\label{alg4}
  \begin{algorithmic}[1]
    \REQUIRE ~~\\
    The state of VMs by defender's observation at time step $t$, $\{O_t(v_1),O_t(v_2),...O_t(v_n)\}$;\\
    A binary $r \times n$-matrix $X_t$;\\
    A binary $u \times n$-matrix $Y_t$;\\
    A binary $q \times n$-matrix $Z_t$;\\
    The number of online users in each VM, $\{\eta_t(v_1),\eta_t(v_2),...\eta_t(v_n)\}$;\\
    \ENSURE ~~\\
    A binary $r \times n$-matrix $X_{t+1}$;\\
    A binary $u \times n$-matrix $Y_{t+1}$;\\
    A binary $q \times n$-matrix $Z_{t+1}$;\\
    \FOR{$O_t(v_i)=1,1 \leqslant i \leqslant n$}
    \IF {$\eta_t(v_i)=0$}
    \STATE Set $x^{t+1}_{i,}=0, y^{t+1}_{i,}=0, z^{t+1}_{i,}=0$;\\
    \ELSE
    \IF {$0 <\eta_t(v_i) \leqslant [\frac{m}{2}]$}
    \STATE Calculate $R^d(v_i),C^d(v_i),R^a(v_i),C^a(v_i)$;\\
    \STATE Find maximum $P^d_{t+1}(v_i)-P^a_{t+1}(v_i)$;\\
    \STATE Set $x^{t+1}_{i,}, y^{t+1}_{i,}, z^{t+1}_{i,}$;\\
    \ELSE
    \STATE Set $z^{t+1}_{i,}=0$;\\
    \STATE Calculate $R^d(v_i),C^d(v_i),R^a(v_i),C^a(v_i)$;\\
    \STATE Find maximum $P^d_{t+1}(v_i)-P^a_{t+1}(v_i)$;\\
    \STATE Set $x^{t+1}_{i,}, y^{t+1}_{i,}$;\\
    \ENDIF
    \ENDIF
    \ENDFOR
    \RETURN all $x^{t+1}_{i,j} \in X_{t+1}, y^{t+1}_{i,j} \in Y_{t+1}, z^{t+1}_{i,j} \in Z_{t+1}$
  \end{algorithmic}
\end{algorithm}

\section{Evaluations and Results}\label{section6}
In this section, we evaluate and analyze the cost-effectiveness and performance of the proposed CES algorithm against DDoS attacks in simulation and experiment. First, we describe the simulation settings and compare our CES algorithm with other existing shuffling algorithms. Then we introduce the experimental settings and implementation of the shuffling scenario in full. Finally, we measure the cost and effectiveness of our proposed CES algorithm in our shuffling scenario with comparisons to a non-shuffling strategy and a random shuffling strategy.

\subsection{Simulation}
\begin{figure*}[htb]
\centering
  \subfigure[]{
    \includegraphics[width=5.65cm]{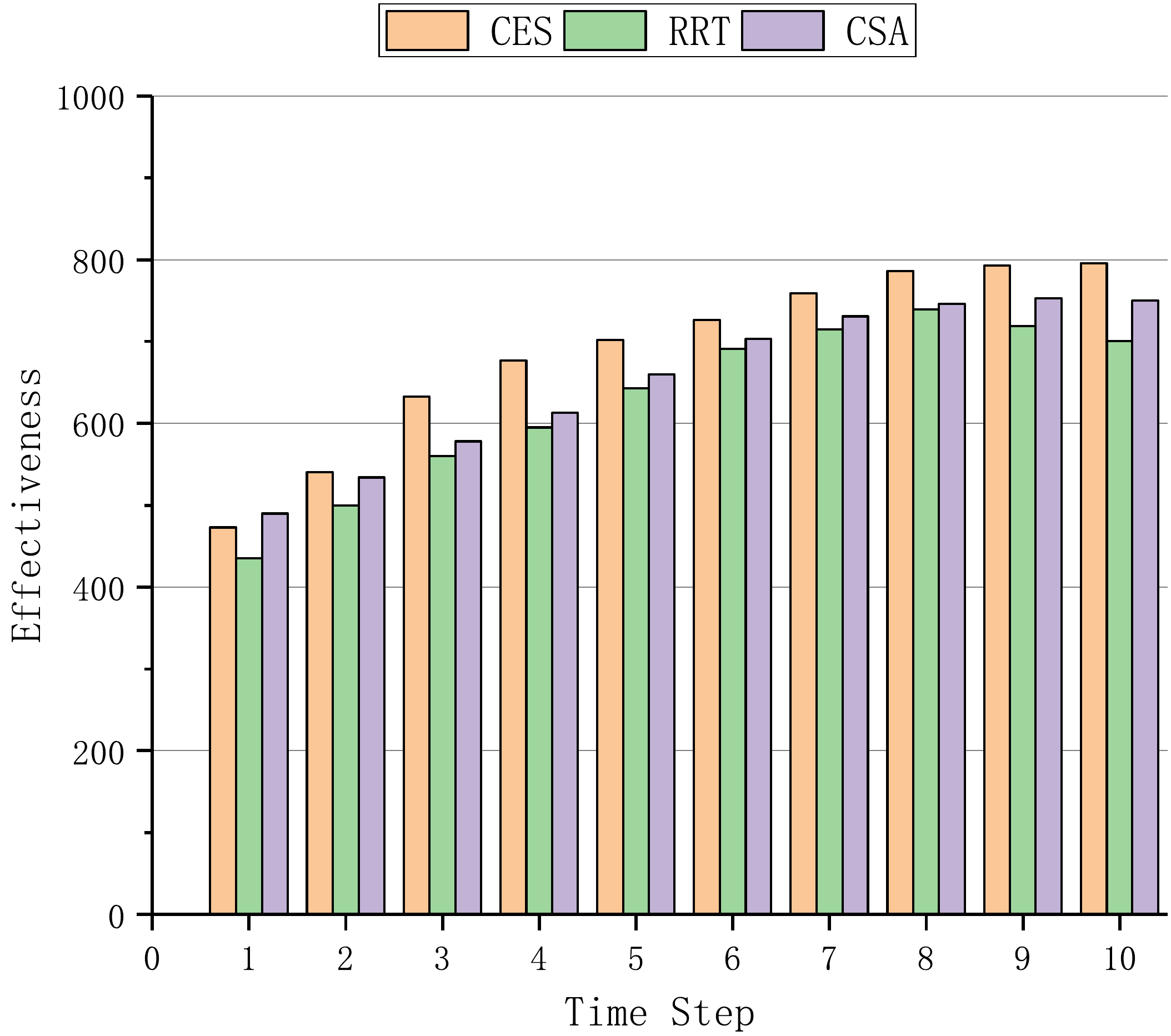}}
  \subfigure[]{
    \includegraphics[width=5.65cm]{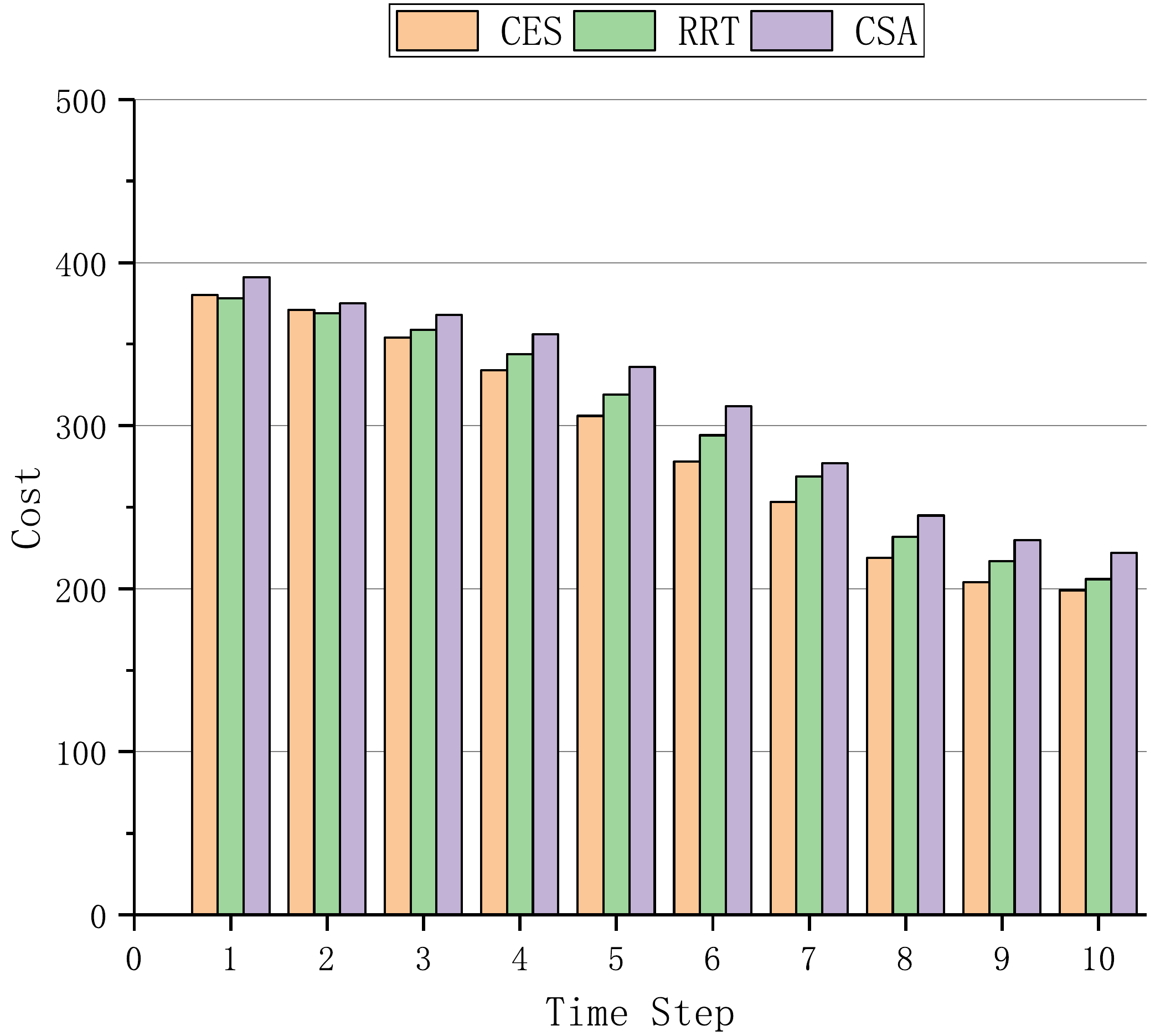}}
      \subfigure[]{
    \includegraphics[width=5.65cm]{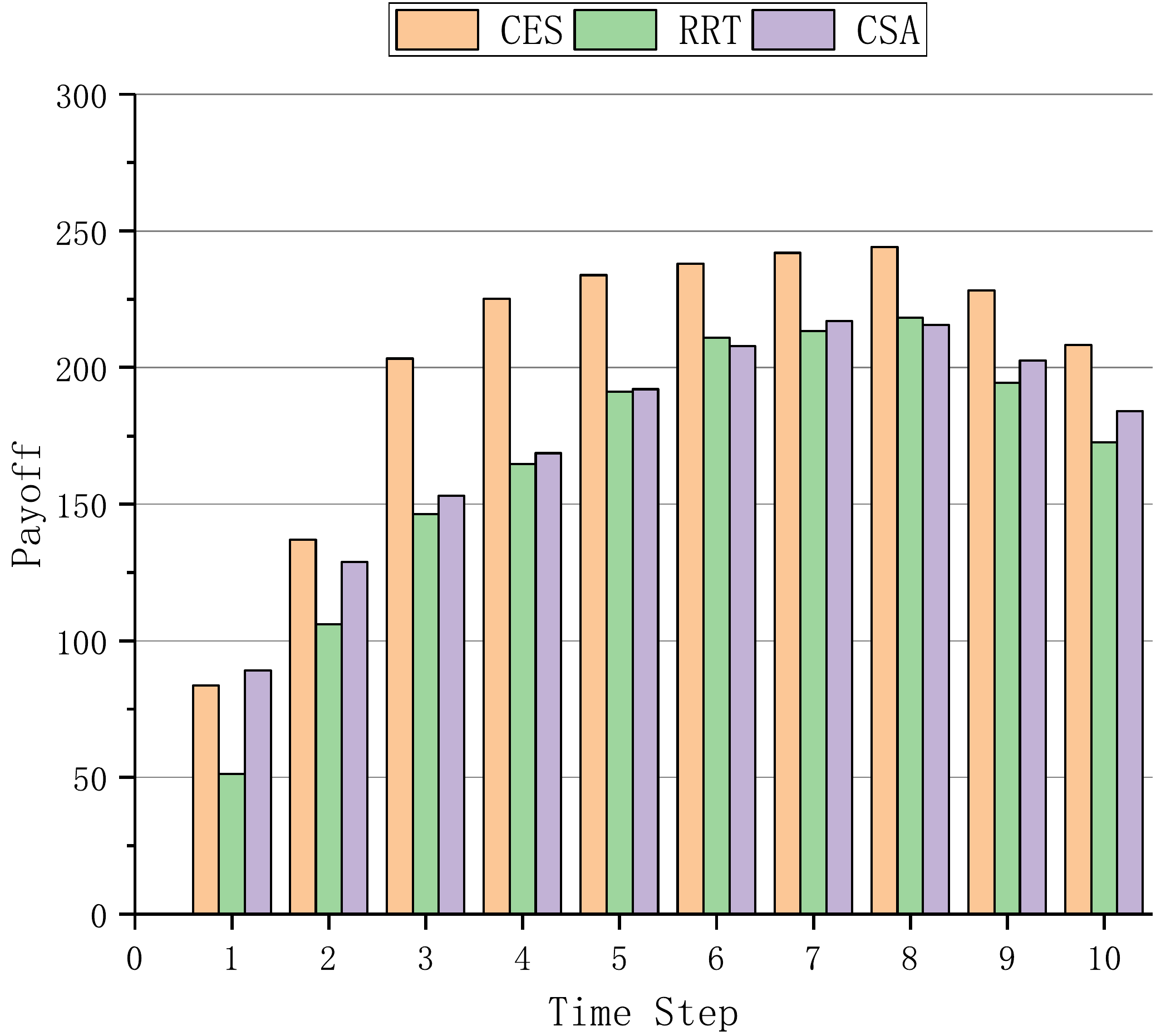}}
  \caption{Comparison of CES,RRT and CSA in the effectiveness, cost and payoff at different time steps.}
  \label{fig1}
\end{figure*}

\begin{figure*}[htb]
\centering
  \subfigure[]{
    \includegraphics[width=5.65cm]{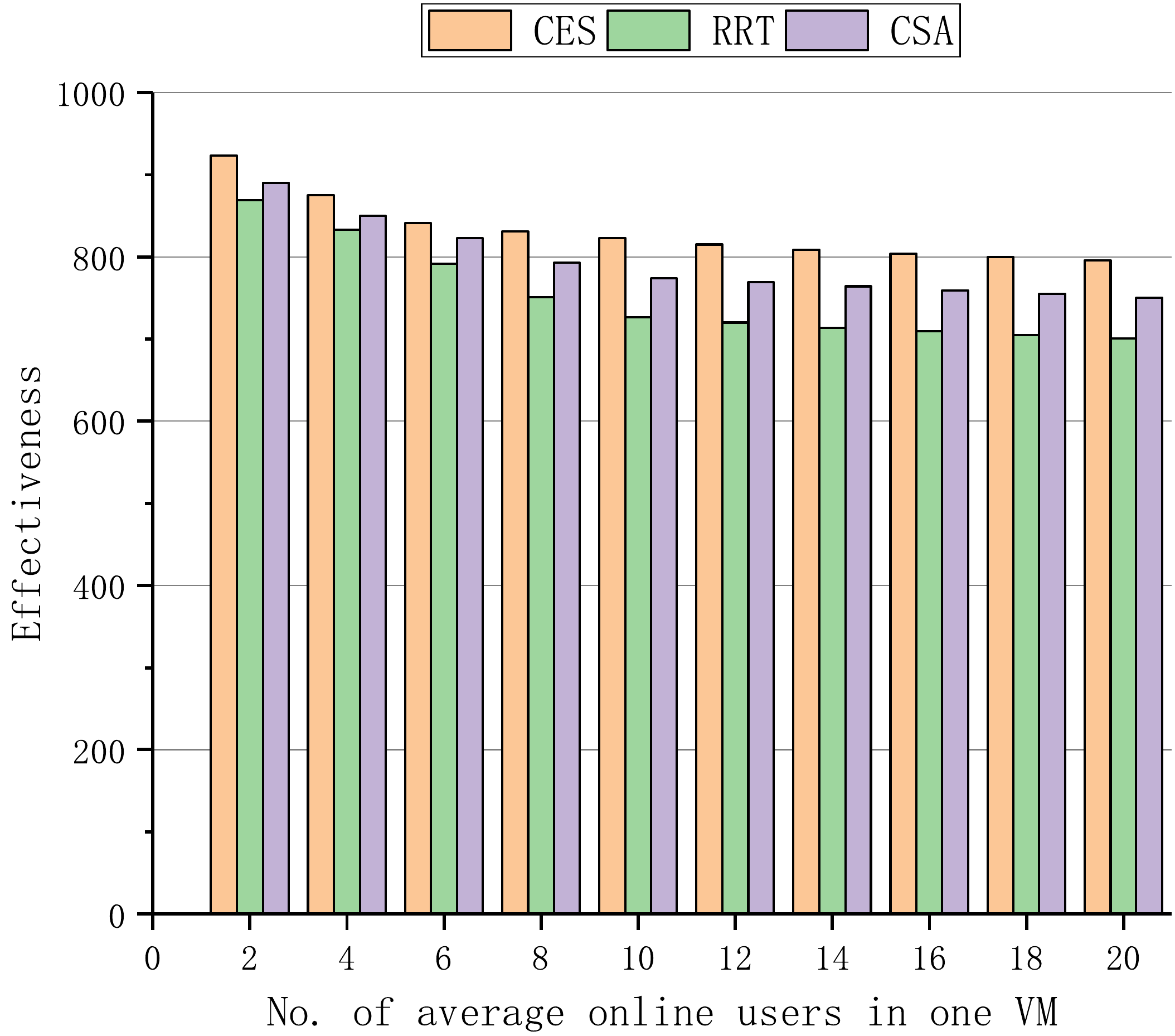}}
  \subfigure[]{
    \includegraphics[width=5.65cm]{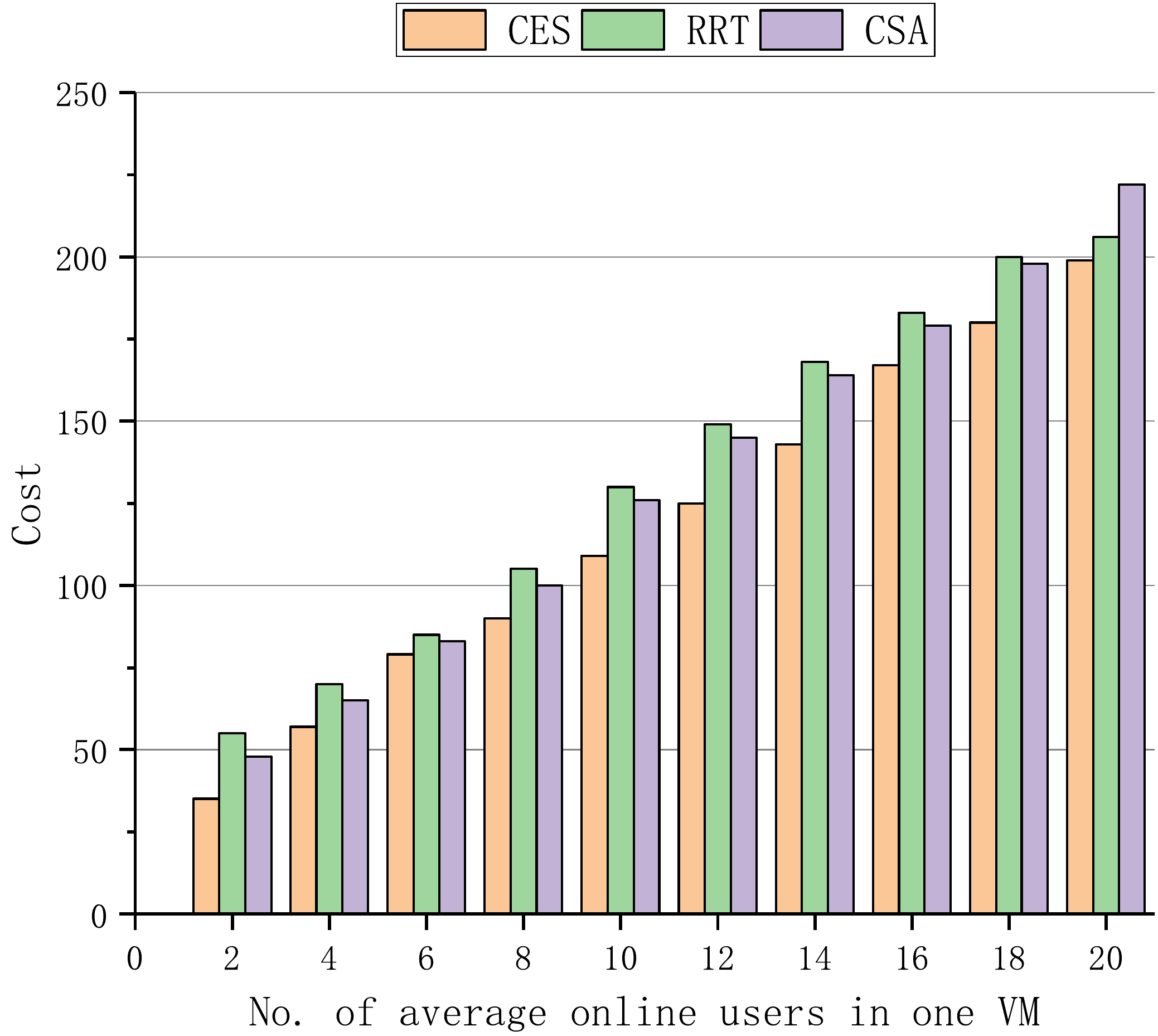}}
      \subfigure[]{
    \includegraphics[width=5.65cm]{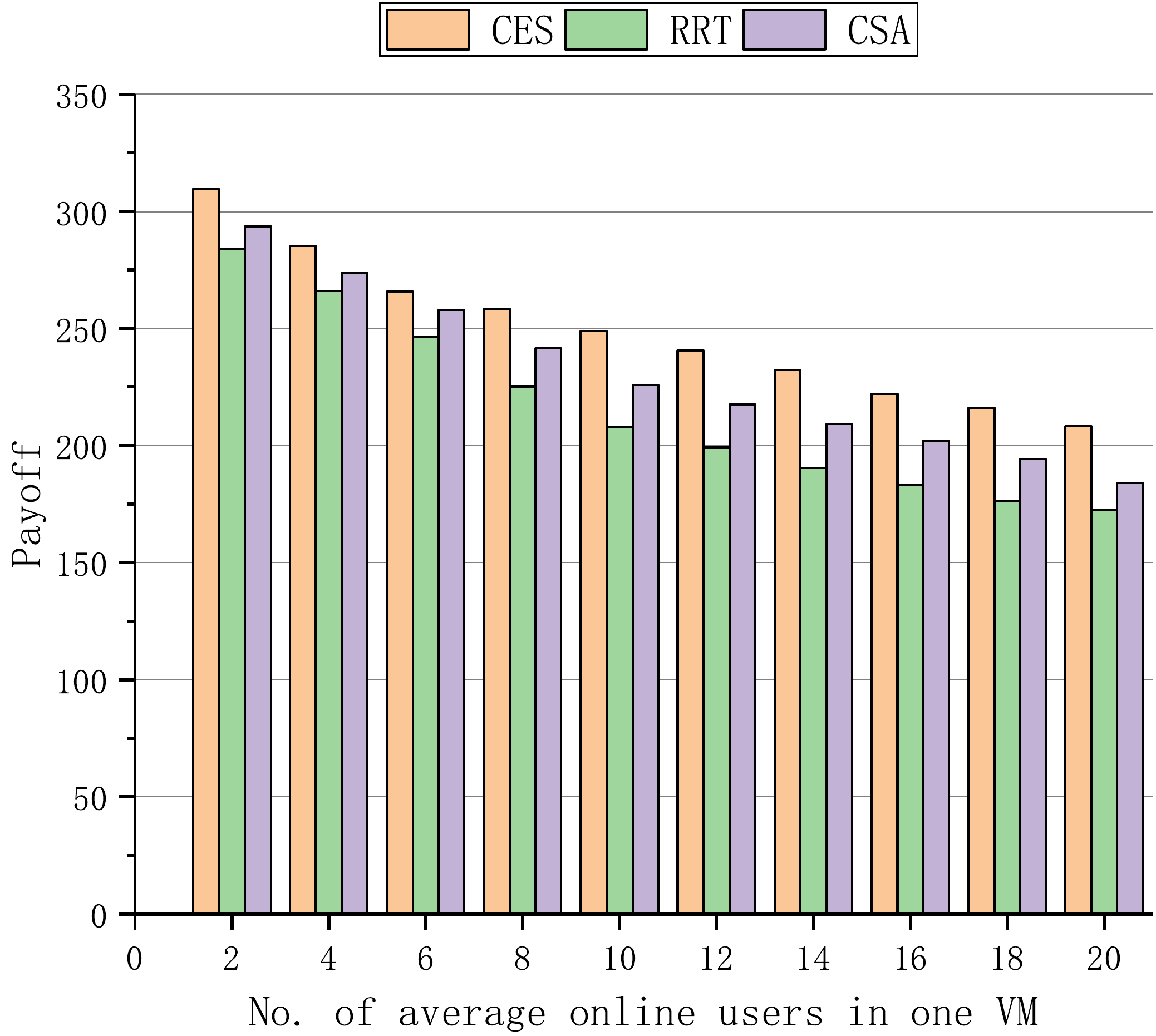}}
  \caption{Comparison of CES,RRT and CSA in the effectiveness, cost and payoff with different number of average online users in one VM.}
  \label{fig2}
\end{figure*}

In the following, we compare the proposed algorithm CES with RRT (Renewal Reward Theory)~\cite{wang2016towards} and CSA (Cost-effective Shuffling Algorithm)~\cite{lin2017cost} to evaluate the effectiveness and cost of shuffling. First, to find out the whole system state transition probability, we implement Algorithm~\ref{alg1} and execute it 10000 times with pre-defined parameters $(m, n, q, r, u)$, where $m=20, n=50, q=1000, r=20, u=100$.

Afterward, we compare the expected value functions of CES with that of RRT and CSA in terms of effectiveness, cost, and payoff. More specifically, the sum cost of single IP hopping, port hopping and VM migration is set to 1, and the effectiveness of successfully defending against an attack is calculated by 1.

Note that RRT is indifferent to the online users of the VMs, and CSA randomly selects half of the users to migrate in a single shuffle. Hence, for a more comprehensive comparison among these algorithms, the parameter $\eta$ is not fixed and ranges from 0 to 20 in the simulation.

Fig. \ref{fig1} and Fig. \ref{fig2} first compare the three algorithms with different time step and different number of average online users in one VM, respectively. In Fig. \ref{fig1}, 1000 users are involved in the shuffling scheme, and the system is allowed to allocate at most 50 VMs for shuffling. In Fig. \ref{fig2}, there are 0 to 20 online users in one VM at time step 10, when the theoretical values of effectiveness, cost and payoff have levelled off. Fig. \ref{fig1} demonstrates that the shuffling approach performs better when the time step increases whereas Fig. \ref{fig2} manifests that the advantage of the shuffling approach decreases when the number of average online users in one VM increases.

Fig. \ref{fig1}(a) and Fig. \ref{fig2}(a) present the effectiveness of shuffling, Fig. \ref{fig1}(b) and Fig. \ref{fig2}(b) show the theoretical cost of shuffling, where the weights for IP hopping, port hopping and migration shuffling mechanisms are set to 0.2, 0.1 and 0.7. The simulation results in Fig. \ref{fig1}(c) and Fig. \ref{fig2}(c) demonstrate that payoff of CES significantly outperforms those of RRT and CSA, where the discount value $\gamma$ is set to 0.9.
\begin{figure*}[t]
\centering
    \includegraphics[width=6.3in]{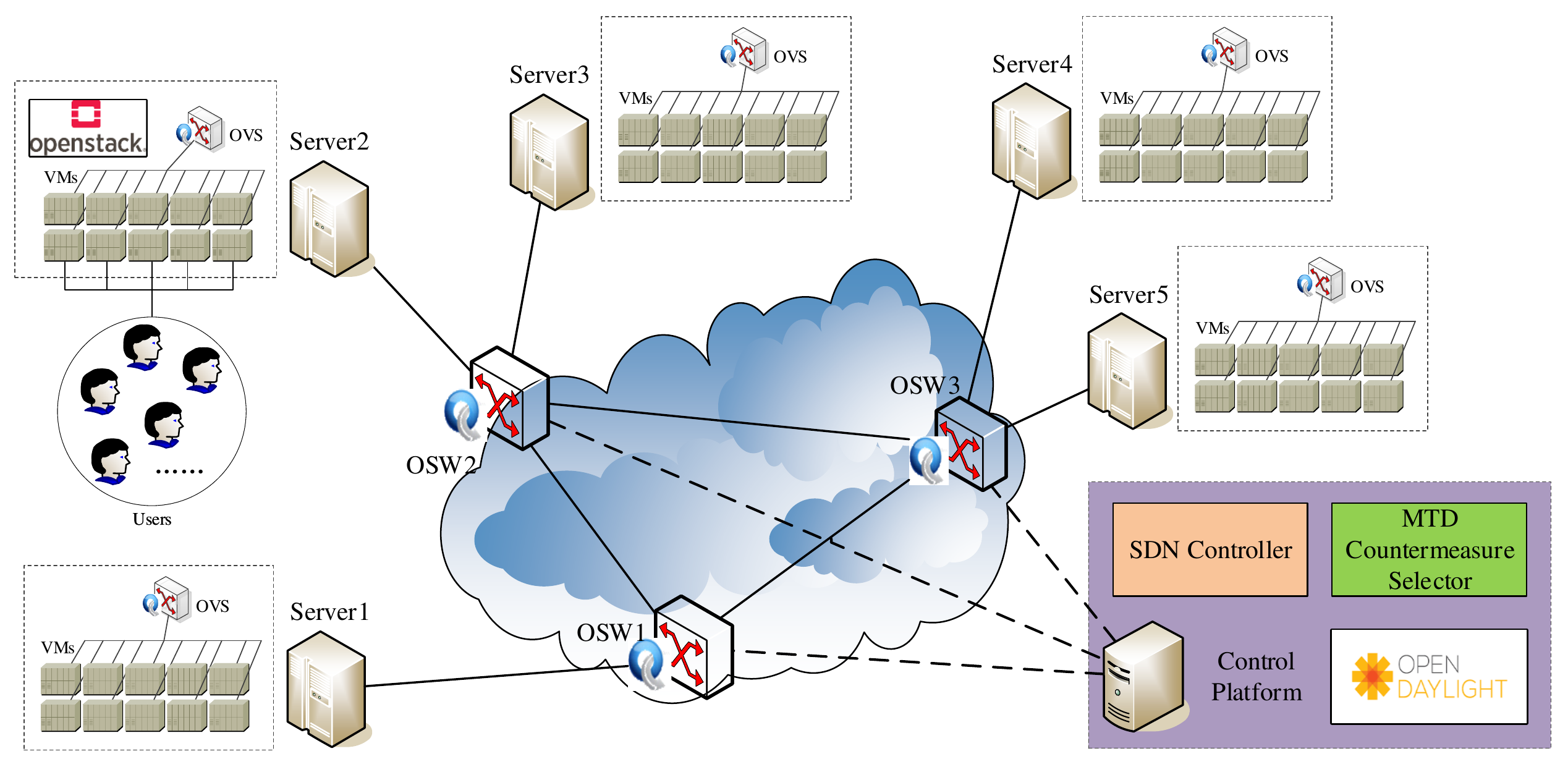}
    \caption{Implementation of the shuffling scenario in an experimental SDN network.}
    \label{fig3}
\end{figure*}

For the theoretical effectiveness of shuffling, Fig. \ref{fig1}(a) and Fig. \ref{fig2}(a) indicate that more effectiveness is gained when the time step increases, and more users lead to less effectiveness for the given time step. For the theoretical shuffling cost, Fig. \ref{fig1}(b) and Fig. \ref{fig2}(b) indicate that the cost linearly decreases when the time step increases and dramatically increases due to the increase of online users. Nevertheless, in consideration of both the effectiveness and cost, our proposed algorithm CES still outperforms RRT and CSA in the payoff, respectively shown in Fig. \ref{fig1}(c) and Fig. \ref{fig2}(c).

The performance of shuffling in CES outperforms shuffling in RRT and CSA due to two reasons. First, the state transition probability of CES fully takes the correlation between states into consideration, while there is no mention of transition probability in RRT, and CSA only represents it as a function without detailed explanations. Second, CES can utilize three kinds of defense mechanism, whereas RRT and CSA can only utilize one. In detail, the underlying reason is because only utilizing VM migrations might introduce more cost if the system state is not so bad, while CES is capable of determining the shuffling mechanism based on the game history and the number of online users.

\subsection{Experimental Settings}
We implement the shuffling scenario in an experimental SDN~\cite{yan2015software} testbed, which is shown in Fig. \ref{fig3}. The testbed that we use for experimental analysis is composed of 5 Dell PowerEdge R720 servers and a Dell PowerEdge R430 server. Each Dell PowerEdge R720 has 32 GB of RAM, 4 TB hard disk storage and 12 core CPU. Dell PowerEdge R430 has 16 GB RAM, 1 TB disk storage and 4 core CPU.

One single server is employed to construct the control platform, using OpenDaylight~\cite{ODL} based SDN controller and PHP Laravel web framework as front-end. For the virtual network deployment, we utilize OpenStack~\cite{Openstack} for computing and network resource provisioning on the other five servers. The VMs are managed and controlled by SDN controller via Open vSwitch~\cite{OpenvSwitch}. A brief description of these existing techniques is given in Section~\ref{ODL}-\ref{OVS}.

In the implementation, we create 50 VMs which are equally allocated to five servers, and each VM is assigned for at most 20 users with equal CPU and memory. In addition, the 50 VMs are organized with different IP and TCP ports, where the attacker can overload the VMs through DDoS attack tools.

\subsubsection{OpenDaylight (ODL)}\label{ODL}
ODL is a open source SDN controller for customizing and automating networks of any size and scale. The OpenDaylight Project arose out of the SDN movement, with a clear focus on network programmability~\cite{ODL}. As a modular and pluggable platform, ODL has the ability to build network functions and services in an adaptable, flexible way.
\subsubsection{OpenStack}\label{Openstack}
OpenStack is a cloud operating system that controls large pools of compute, storage, and networking resources throughout a datacenter, it provides a virtual layer on physical servers, decoupling underlying hardware from the workload. All resources can be managed through a dashboard that gives administrators control while empowering their users to provision resources through a web interface~\cite{Openstack}.

\subsubsection{Open vSwitch (OVS)}\label{OVS}
The most popular virtual switch implementation OVS is heavily used in cloud computing frameworks. It is designed to enable massive network automation through programmatic extension, while still supporting standard management interfaces and protocols (e.g. NetFlow, sFlow, IPFIX, RSPAN, CLI, LACP, 802.1ag)~\cite{OpenvSwitch}.

\subsection{Results}
First, we implement our CES algorithm and execute the program as an application on the control platform. Then, we compare our cost-effective shuffling method with non-shuffling and random shuffling scenario in terms of overhead and performance.
\subsubsection{Overhead of SDN Controller's CPU Load}
\begin{figure}[b]
\centering
    \includegraphics[width=2.5in]{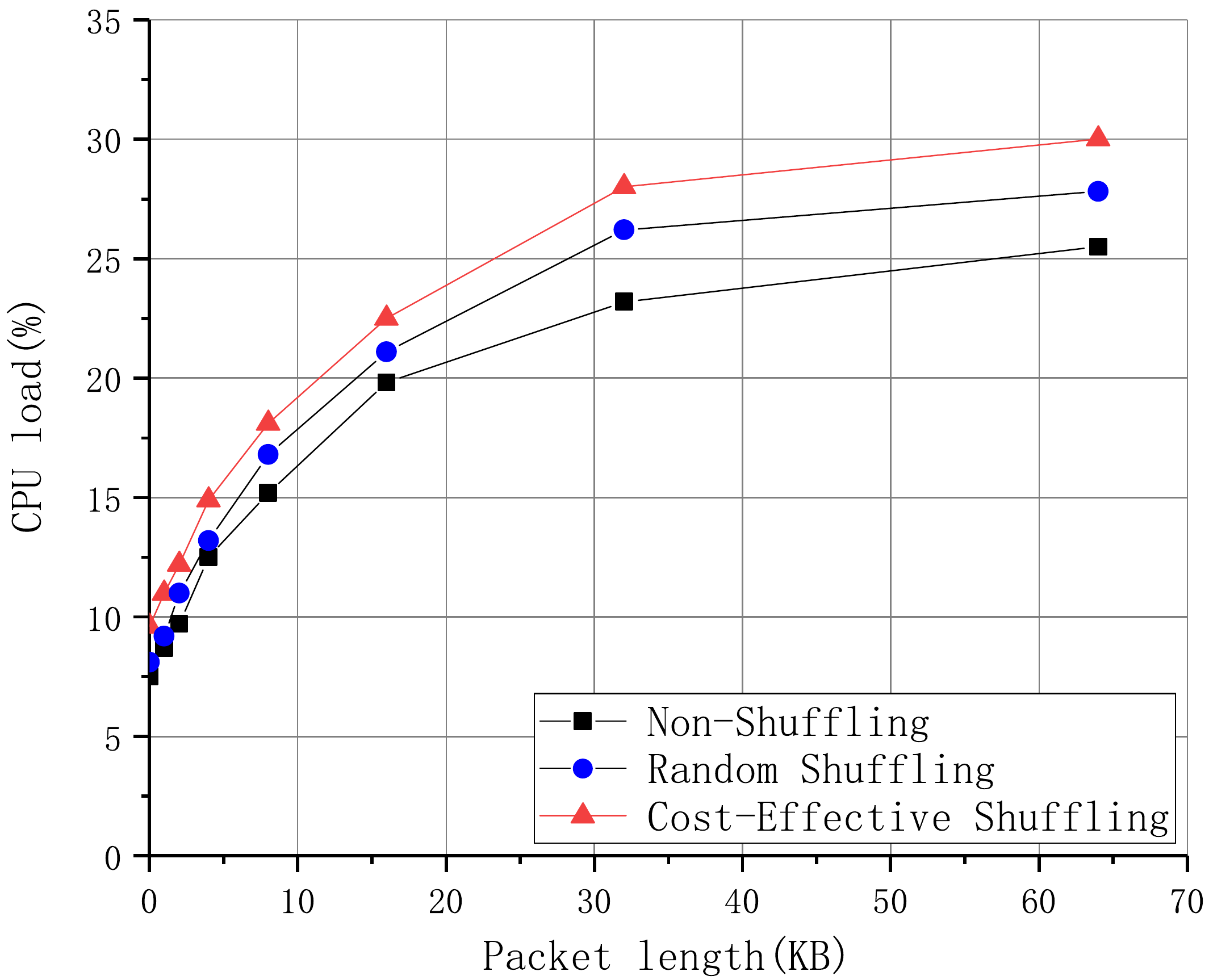}
    \caption{CPU load of SDN controller in different shuffling scenarios.}
    \label{fig4}
\end{figure}

In order to evaluate the processing overhead on the SDN controller consumed by the shuffling scenario, we use packets of different lengths to communicate and evaluate the influence to SDN controller's CPU load among different scenarios, which is shown in Fig. \ref{fig4}. The extra CPU load is about 2.1 $\%$-4.8 $\%$ compared to non-shuffling scenario and about 1.2 $\%$-2.2 $\%$ compared to random shuffling scenario. The extra processing overhead on the SDN controller is not heavy, and is in an acceptable level when CES has been deployed.

\subsubsection{Overhead of the Shuffling Process}

\begin{figure}[b]
\centering
    \includegraphics[width=2.5in]{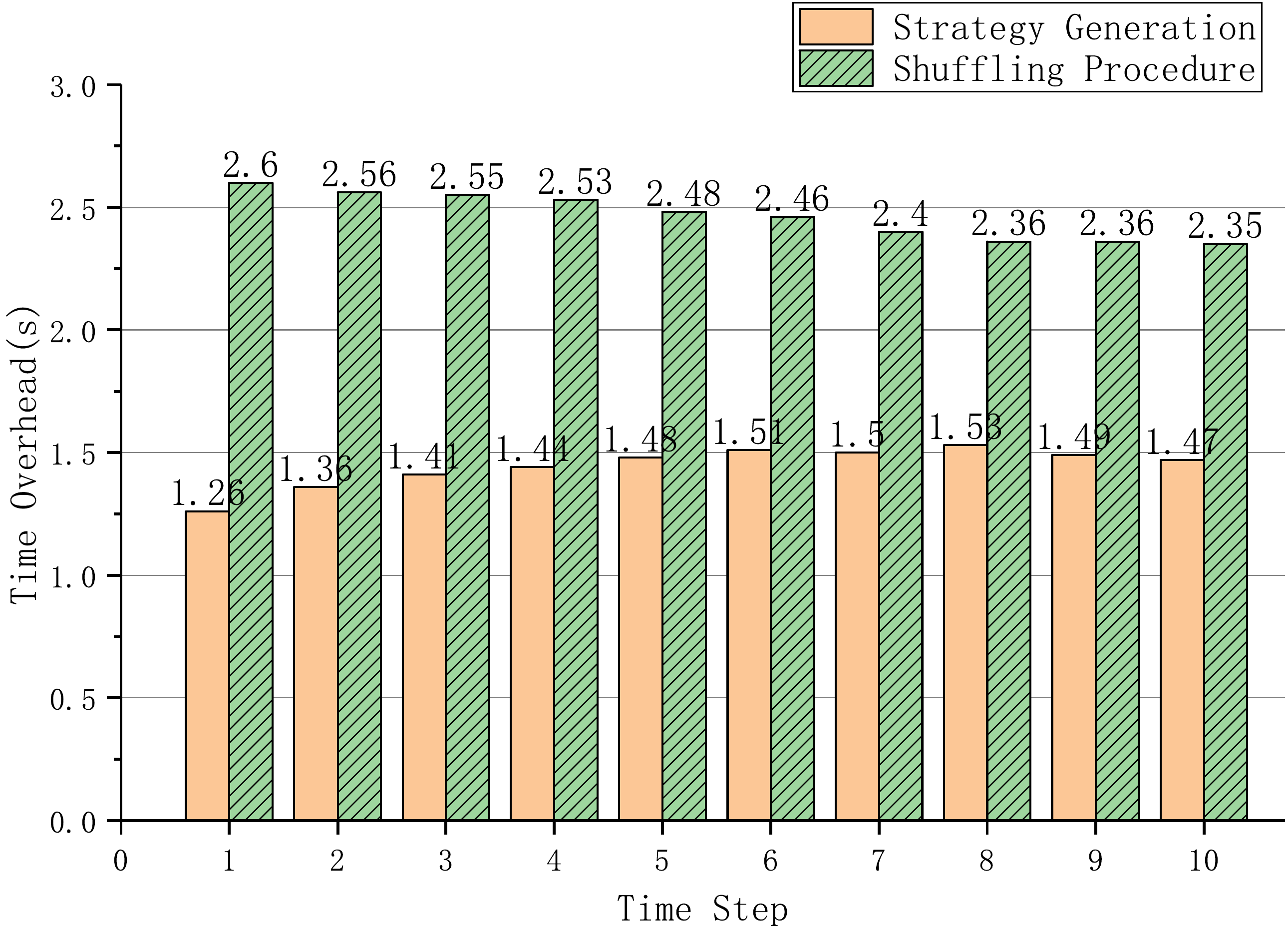}
    \caption{Time consumption of strategy generation and shuffling procedure in each shuffle.}
    \label{fig5}
\end{figure}
In order to evaluate the overhead of each shuffle consumed, we combine the time consumption of defense strategy generation (the running time of CES) and shuffling procedure to represent the overhead of the whole shuffling process, which is shown in Fig. \ref{fig5}.

In general, the results indicate that our approach in total requires 3.82-3.97s in each shuffle, including the time consumption of defense strategy generation and shuffling procedure. This is an acceptable time for users to wait during the restart of services. The time consumption of defense strategy generation increases when the time step increases. However, as the time step increases, there is a slight decrease on the time consumption of shuffling procedure.

\begin{figure}[b]
\centering
    \includegraphics[width=2.5in]{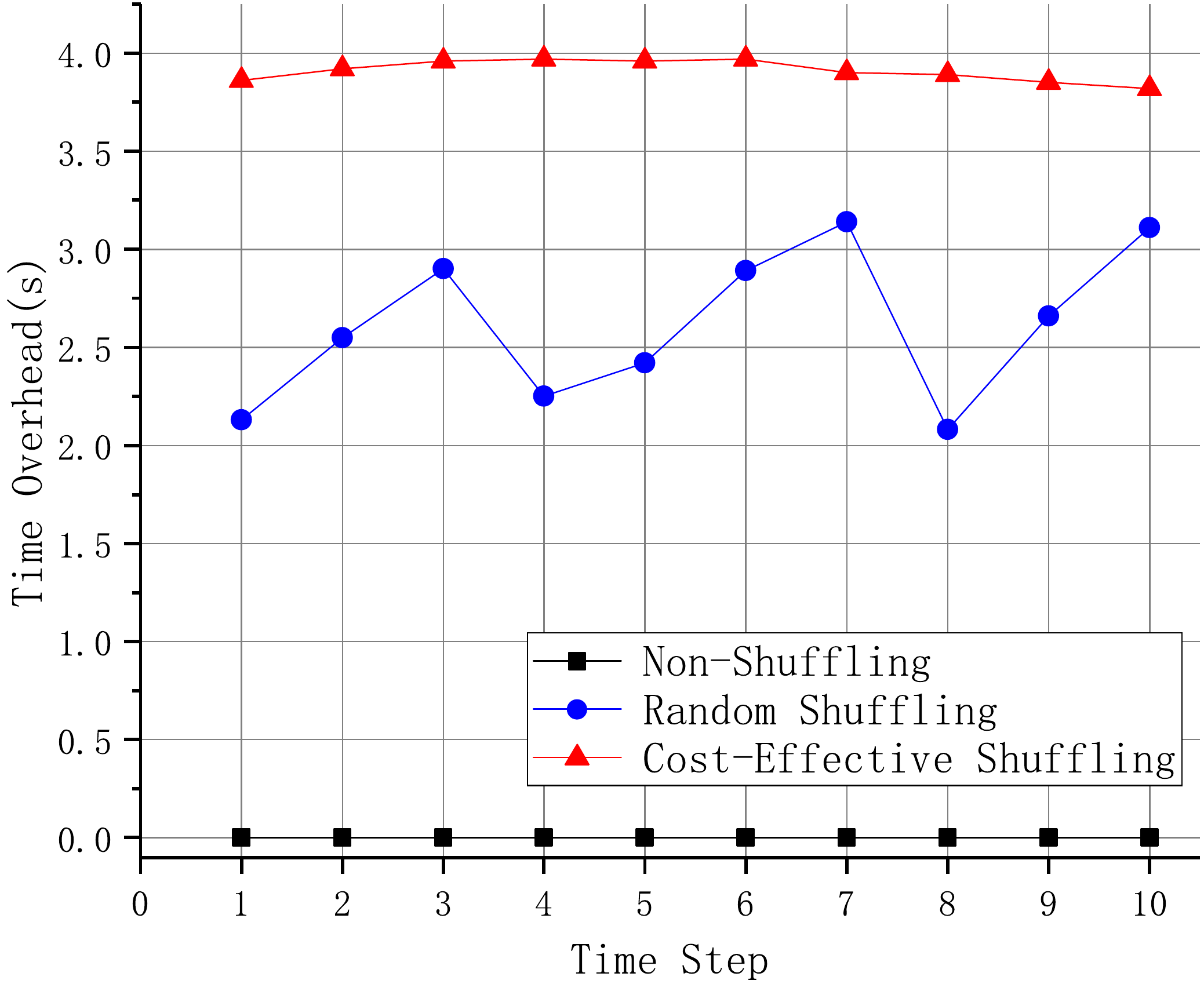}
    \caption{Comparison of time overhead in different shuffling scenarios.}
    \label{fig6}
\end{figure}

In addition, we compare the time overhead among three shuffling scenarios. As seen in Fig. \ref{fig6}, non-shuffling method has no time overhead and random shuffling spends 2.08-3.14s at each time step. Though more time was consumed using the proposed cost-effective shuffling method due to its defense strategy generation, it was still able to keep the time-cost within 5 seconds.

\subsubsection{Performance of Resisting DDoS Attacks}

\begin{figure}[b]
\centering
    \includegraphics[width=2.5in]{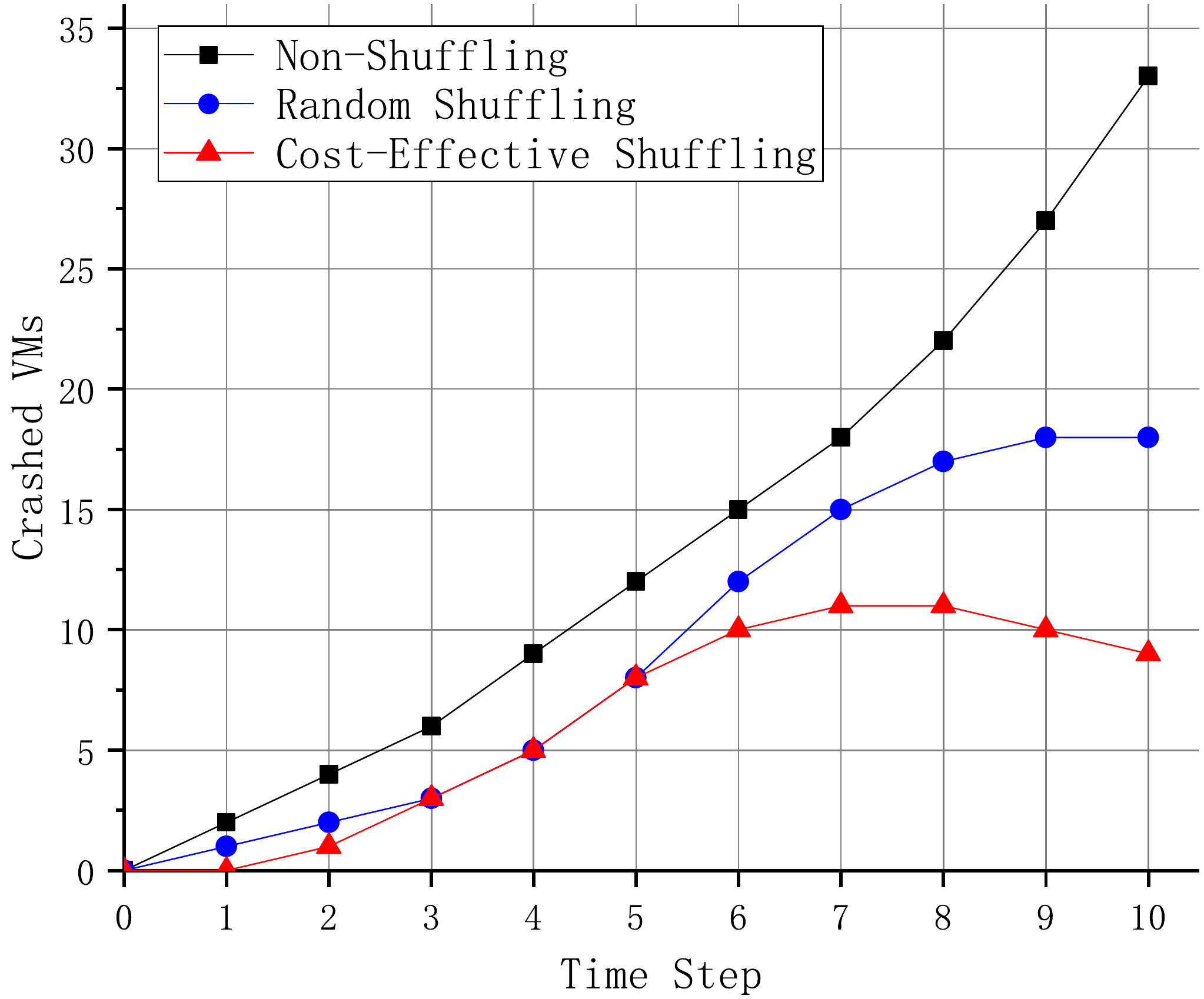}
    \caption{Numbers of crashed VMs in different shuffling scenarios.}
    \label{fig7}
\end{figure}
Finally, to evaluate the capability of our proposed method to resist DDoS attacks, we construct a typical SYN (synchronize) flood DDoS attack tool using hping3~\cite{rghanti2015design} and carry out DDoS attacks on the protected VMs one by one in our shuffling scenario. Test results are shown in Fig. \ref{fig7}.

It is obvious that random and cost-effective shuffling methods have a better performance than non-shuffling method in the ability against DDoS attacks. It can be also seen from Fig. \ref{fig7} that when suffering from DDoS attacks, non-shuffling and random shuffling methods were outperformed by our proposed cost-effective shuffling method. CES was faster in recovering protected systems, better at keeping system services online, and better at restricting the number of crashed VMs within the limited time steps.
\section{Conclusions}\label{section7}
This paper proposed a cost-effective method in the shuffling-based moving target defense scenario against DDoS attacks. First, we described a threat model to characterize the behavior of attackers and defense mechanism. The interaction was modeled between the attacker and the defender with Multi-Objective Markov Decision Processes, and the game process at each time step was described in detail with game payoff and strategy. Then, in order to maximize the defender's payoff and minimize the attacker's payoff, the CES algorithm was designed to seek the best trade-off between shuffling cost and effectiveness in the proposed scenario.

The cost-effectiveness of the proposed CES algorithm was evaluated in simulation and outperformed other existing algorithms, such as RRT and CSA. In addition, CES was deployed on an SDN based shuffling scenario and evaluated in terms of overhead and performance. The comparison with a non-shuffling approach and a random shuffling approach showed several key advantages of the proposed algorithm. First, the lower required CPU and time overhead ensured the feasibility of the proposed method. Second, it is evident that the deployment of CES was beneficial for improving overall system security and for protecting the system against DDoS attacks effectively.

The next step is to introduce other MTD technologies (such as service hopping, path hopping, etc.) into defense mechanism and fine tune the quantitative analysis of our research. In addition, making full use of the characteristics of game theory, multi-stage game between the attacker and the defender will be further studied.

\begin{acks}
This work was supported by National Key R$\&$D Program of China (2018YFB1800602, 2017YFB0801703), National Natural Science Foundation of China (61602114), and CERNET Innovation Project (NGIICS20190101, NGII20170406).
\end{acks}

%
\bibliographystyle{unsrt}
\bibliographystyle{ACM-Reference-Format}
\bibliography{acmart}

\begin{thebibliography}{10}

\bibitem{carvalho2014moving}
Marco Carvalho and Richard Ford.
\newblock Moving-target defenses for computer networks.
\newblock {\em IEEE Security \& Privacy}, 12(2):73--76, 2014.

\bibitem{cai2016moving}
Gui-lin Cai, Bao-sheng Wang, Wei Hu, and Tian-zuo Wang.
\newblock Moving target defense: state of the art and characteristics.
\newblock {\em Frontiers of Information Technology \& Electronic Engineering},
  17(11):1122--1153, 2016.

\bibitem{lei2018moving}
Cheng Lei, Hong-Qi Zhang, Jing-Lei Tan, Yu-Chen Zhang, and Xiao-Hu Liu.
\newblock Moving target defense techniques: A survey.
\newblock {\em Security and Communication Networks}, 2018, 2018.

\bibitem{manadhata2004measuring}
Pratyusa Manadhata and Jeannette~M Wing.
\newblock Measuring a system's attack surface.
\newblock Technical report, CARNEGIE-MELLON UNIV PITTSBURGH PA SCHOOL OF
  COMPUTER SCIENCE, 2004.

\bibitem{zangeneh2018cost}
Vahid Zangeneh and Mehdi Shajari.
\newblock A cost-sensitive move selection strategy for moving target defense.
\newblock {\em Computers \& Security}, 75:72--91, 2018.

\bibitem{roijers2017multi}
Diederik~M Roijers and Shimon Whiteson.
\newblock Multi-objective decision making.
\newblock {\em Synthesis Lectures on Artificial Intelligence and Machine
  Learning}, 11(1):1--129, 2017.

\bibitem{pal2013managed}
Partha Pal, Richard Schantz, Aaron Paulos, and Brett Benyo.
\newblock Managed execution environment as a moving-target defense
  infrastructure.
\newblock {\em IEEE Security \& Privacy}, 12(2):51--59, 2013.

\bibitem{rahman2014moving}
Mohammad~Ashiqur Rahman, Ehab Al-Shaer, and Rakesh~B Bobba.
\newblock Moving target defense for hardening the security of the power system
  state estimation.
\newblock In {\em Proceedings of the First ACM Workshop on Moving Target
  Defense}, pages 59--68. ACM, 2014.

\bibitem{gillani2015agile}
Fida Gillani, Ehab Al-Shaer, Samantha Lo, Qi~Duan, Mostafa Ammar, and Ellen
  Zegura.
\newblock Agile virtualized infrastructure to proactively defend against cyber
  attacks.
\newblock In {\em 2015 IEEE Conference on Computer Communications (INFOCOM)},
  pages 729--737. IEEE, 2015.

\bibitem{chang2018fast}
Sang-Yoon Chang, Younghee Park, and Bhavana Babu~Ashok Babu.
\newblock Fast ip hopping randomization to secure hop-by-hop access in sdn.
\newblock {\em IEEE Transactions on Network and Service Management},
  16(1):308--320, 2018.

\bibitem{carroll2014analysis}
Thomas~E Carroll, Michael Crouse, Errin~W Fulp, and Kenneth~S Berenhaut.
\newblock Analysis of network address shuffling as a moving target defense.
\newblock In {\em 2014 IEEE International Conference on Communications (ICC)},
  pages 701--706. IEEE, 2014.

\bibitem{crouse2015probabilistic}
Michael Crouse, Bryan Prosser, and Errin~W Fulp.
\newblock Probabilistic performance analysis of moving target and deception
  reconnaissance defenses.
\newblock In {\em Proceedings of the Second ACM Workshop on Moving Target
  Defense}, pages 21--29. ACM, 2015.

\bibitem{zhang2017network}
Hong-qi Zhang, Cheng Lei, De-xian Chang, and Ying-jie Yang.
\newblock Network moving target defense technique based on collaborative
  mutation.
\newblock {\em computers \& security}, 70:51--71, 2017.

\bibitem{kampanakis2014sdn}
Panos Kampanakis, Harry Perros, and Tsegereda Beyene.
\newblock Sdn-based solutions for moving target defense network protection.
\newblock In {\em Proceeding of IEEE International Symposium on a World of
  Wireless, Mobile and Multimedia Networks 2014}, pages 1--6. IEEE, 2014.

\bibitem{huang2011introducing}
Yih Huang and Anup~K Ghosh.
\newblock Introducing diversity and uncertainty to create moving attack
  surfaces for web services.
\newblock In {\em Moving target defense}, pages 131--151. Springer, 2011.

\bibitem{kansal2017ddos}
Vaishali Kansal and Mayank Dave.
\newblock Ddos attack isolation using moving target defense.
\newblock In {\em 2017 International Conference on Computing, Communication and
  Automation (ICCCA)}, pages 511--514. IEEE, 2017.

\bibitem{alavizadeh2018evaluation}
Hooman Alavizadeh, Julian Jang-Jaccard, and Dong~Seong Kim.
\newblock Evaluation for combination of shuffle and diversity on moving target
  defense strategy for cloud computing.
\newblock In {\em 2018 17th IEEE International Conference On Trust, Security
  And Privacy In Computing And Communications/12th IEEE International
  Conference On Big Data Science And Engineering (TrustCom/BigDataSE)}, pages
  573--578. IEEE, 2018.

\bibitem{hong2015assessing}
Jin~B Hong and Dong~Seong Kim.
\newblock Assessing the effectiveness of moving target defenses using security
  models.
\newblock {\em IEEE Transactions on Dependable and Secure Computing},
  13(2):163--177, 2015.

\bibitem{bopche2017graph}
Ghanshyam~S Bopche and Babu~M Mehtre.
\newblock Graph similarity metrics for assessing temporal changes in attack
  surface of dynamic networks.
\newblock {\em Computers \& Security}, 64:16--43, 2017.

\bibitem{hong2018dynamic}
Jin~B Hong, Simon~Yusuf Enoch, Dong~Seong Kim, Armstrong Nhlabatsi, Noora
  Fetais, and Khaled~M Khan.
\newblock Dynamic security metrics for measuring the effectiveness of moving
  target defense techniques.
\newblock {\em Computers \& Security}, 79:33--52, 2018.

\bibitem{xiong2019effectiveness}
Xin-Li Xiong, Lin Yang, and Guang-Sheng Zhao.
\newblock Effectiveness evaluation model of moving target defense based on
  system attack surface.
\newblock {\em IEEE Access}, 7:9998--10014, 2019.

\bibitem{zhang2019efficient}
Huan Zhang, Kangfeng Zheng, Xiujuan Wang, Shoushan Luo, and Bin Wu.
\newblock Efficient strategy selection for moving target defense under multiple
  attacks.
\newblock {\em IEEE Access}, 2019.

\bibitem{prakash2015empirical}
Achintya Prakash and Michael~P Wellman.
\newblock Empirical game-theoretic analysis for moving target defense.
\newblock In {\em Proceedings of the Second ACM Workshop on Moving Target
  Defense}, pages 57--65. ACM, 2015.

\bibitem{feng2017signaling}
Xiaotao Feng, Zizhan Zheng, Derya Cansever, Ananthram Swami, and Prasant
  Mohapatra.
\newblock A signaling game model for moving target defense.
\newblock In {\em IEEE INFOCOM 2017-IEEE Conference on Computer
  Communications}, pages 1--9. IEEE, 2017.

\bibitem{miehling2015optimal}
Erik Miehling, Mohammad Rasouli, and Demosthenis Teneketzis.
\newblock Optimal defense policies for partially observable spreading processes
  on bayesian attack graphs.
\newblock In {\em Proceedings of the Second ACM Workshop on Moving Target
  Defense}, pages 67--76. ACM, 2015.

\bibitem{hu2017online}
Zhisheng Hu, Minghui Zhu, and Peng Liu.
\newblock Online algorithms for adaptive cyber defense on bayesian attack
  graphs.
\newblock In {\em MTD@ CCS}, pages 99--109, 2017.

\bibitem{zheng2018markov}
Jianjun Zheng and Akbar Siami~Namin.
\newblock A markov decision process to determine optimal policies in moving
  target.
\newblock In {\em Proceedings of the 2018 ACM SIGSAC Conference on Computer and
  Communications Security}, pages 2321--2323. ACM, 2018.

\bibitem{lei2018incomplete}
Cheng Lei, Hong-Qi Zhang, Li-Ming Wan, Lu~Liu, and Duo-he Ma.
\newblock Incomplete information markov game theoretic approach to strategy
  generation for moving target defense.
\newblock {\em Computer Communications}, 116:184--199, 2018.

\bibitem{lin2017cost}
Yi-Hui Lin, Jian-Jhih Kuo, De-Nian Yang, and Wen-Tsuen Chen.
\newblock A cost-effective shuffling-based defense against http ddos attacks
  with sdn/nfv.
\newblock In {\em 2017 IEEE International Conference on Communications (ICC)},
  pages 1--7. IEEE, 2017.

\bibitem{wang2016towards}
Huangxin Wang, Fei Li, and Songqing Chen.
\newblock Towards cost-effective moving target defense against ddos and covert
  channel attacks.
\newblock In {\em Proceedings of the 2016 ACM Workshop on Moving Target
  Defense}, pages 15--25. ACM, 2016.

\bibitem{yan2015software}
Qiao Yan, F~Richard Yu, Qingxiang Gong, and Jianqiang Li.
\newblock Software-defined networking (sdn) and distributed denial of service
  (ddos) attacks in cloud computing environments: A survey, some research
  issues, and challenges.
\newblock {\em IEEE communications surveys \& tutorials}, 18(1):602--622, 2015.

\bibitem{ODL}
Open{D}ay{L}ight, 2019.
\newblock Home - OpenDaylight. \url{https://www.opendaylight.org/}.

\bibitem{Openstack}
Open{S}tack, 2019.
\newblock Build the future of Open Infrastructure.
  \url{https://www.openstack.org/}.

\bibitem{OpenvSwitch}
Open v{S}witch, 2019.
\newblock Open vSwitch. \url{http://www.openvswitch.org/}.

\bibitem{rghanti2015design}
Shaila RGhanti and GM~GM~Naik.
\newblock Design of system on chip for generating syn flood attack to test the
  performance of the security system.
\newblock {\em International Journal of Computer Applications}, 122(7):14--17,
  2015.

\end{thebibliography}

%
\appendix
\end{document}